\documentclass{stacs_proc}
\usepackage{amssymb}
\usepackage{graphicx}
\usepackage{psfrag}
\usepackage{url}

\begin{document}
\title[Shortest Vertex-Disjoint Two-Face Paths in Planar Graphs]%
{Shortest Vertex-Disjoint Two-Face Paths\\ in Planar Graphs}

\author[ens]{\'Eric Colin de Verdi\`ere}{\'Eric Colin de Verdi\`ere}
\address[ens]{Laboratoire d'informatique
  \newline \'Ecole normale sup\'erieure, CNRS
  \newline 45, rue d'Ulm
  \newline 75005 Paris
  \newline France
}
\email{Eric.Colin.de.Verdiere@ens.fr}
\urladdr{http://www.di.ens.fr/\~{}colin/}

\author[cwi]{Alexander Schrijver}{Alexander Schrijver}
\address[cwi]{Centrum voor Wiskunde en Informatica
  \newline Kruislaan 413
  \newline 1098 SJ Amsterdam
  \newline Netherlands
}
\email{lex@cwi.nl}
\urladdr{http://homepages.cwi.nl/\~{}lex/}

\thanks{Most of this work was done while the first author was
  visiting the second author at CWI Amsterdam.}



\begin{abstract}
  Let $G$ be a directed planar graph of complexity~$n$, each arc having a
  nonnegative length.  Let $s$ and~$t$ be two distinct faces of~$G$; let
  $s_1,\ldots,s_k$ be vertices incident with~$s$; let $t_1,\ldots,t_k$ be
  vertices incident with~$t$.  We give an algorithm to compute $k$ pairwise
  vertex-disjoint paths connecting the pairs $(s_i,t_i)$ in~$G$, with
  minimal total length, in $O(kn\log n)$ time.
\end{abstract}

\maketitle

\stacsheading{2008}{181-192}{Bordeaux}
\firstpageno{181}

\newcommand{\RR}{\ensuremath{\mathbf{R}}}
\newcommand{\ZZ}{\ensuremath{\mathbf{Z}}}
\newcommand{\transp}{\ensuremath{^{\top}}}
\newcommand{\ptransp}{\ensuremath{^{\prime\top}}}
\newcommand{\inv}{\ensuremath{^{-1}}}
\newcommand{\source}{\ensuremath{\mathrm{source}}}
\newcommand{\target}{\ensuremath{\mathrm{target}}}

\section{Introduction}

The \emph{vertex-disjoint paths problem} is described as follows: given any
(directed or undirected) graph and $k$ pairs $(s_1,t_1),\ldots,(s_k,t_k)$
of vertices, find $k$ pairwise vertex-disjoint paths connecting the pairs
$(s_i,t_i)$, if they exist.  This problem is well-known also because of its
motivation by VLSI-design.

For a fixed number $k$ of pairs of terminals, this problem is
polynomial-time solvable in a directed planar graph, as shown by
Schrijver~\cite{s-fdpdp-94}, and in any undirected graph, as shown by
Robertson and Seymour~\cite{rs-gmdpp-95}.  However,
Raghavan~\cite{r-rrdhs-86} and Kramer and van Leeuwen~\cite{kl-cwrfm-84}
proved that it is NP-hard when $k$ is not fixed, even on a planar
undirected graph; it belongs to the more general class of \emph{integer
  multicommodity flow problems}~\cite[Chapter 70]{s-cope-03}, many variants
of which are NP-hard.

If the graph is planar, two special cases are solvable in time linear in
the complexity of the graph, even if~$k$ is not fixed:
\begin{enumerate}\renewcommand{\theenumi}{\alph{enumi}}
\item\label{enum_prob_ii} if all terminals lie on the outer face, as proved
  by Suzuki et al.~\cite{san-fsfpg-90};
\item\label{enum_prob_i} if the terminals $s_1,\ldots,s_k$ are incident
  with a common face~$s$, the terminals $t_1,\ldots,t_k$ are incident with
  a common face~$t$, and the faces~$s$ and~$t$ are distinct, as proved by
  Ripphausen-Lipa et al.~\cite{rww-ltadt-96}.
\end{enumerate}

\begin{figure}
  \centerline{%
    \psfrag{a1}{$s_1$}\psfrag{a2}{$s_2$}\psfrag{a3}{$s_3$}\psfrag{a4}{$s_4$}%
    \psfrag{b1}{$t_1$}\psfrag{b2}{$t_2$}\psfrag{b3}{$t_3$}\psfrag{b4}{$t_4$}%
    \includegraphics[width=.45\linewidth]{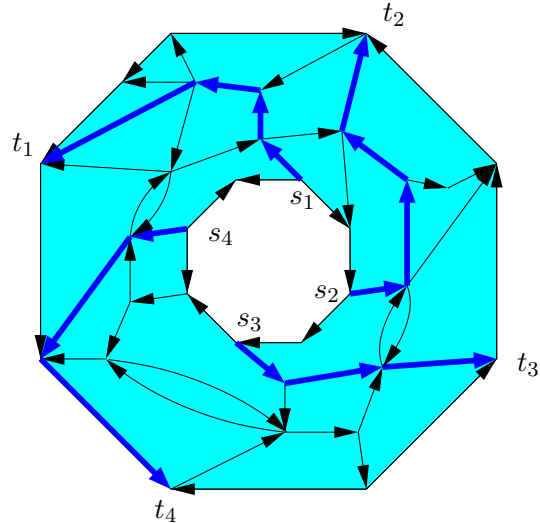}}
  \caption{An instance of the problem and a solution (in bold lines).}
  \label{fig_problem}
\end{figure}

In this paper, we consider a graph where each edge has a nonnegative
length, and we wish to solve the vertex-disjoint paths problem using paths
with minimal total length.  Of course, this is harder than the
vertex-disjoint paths problem.  In case~(\ref{enum_prob_ii}), the problem
is known to be solvable in polynomial time (even if $k$ is not fixed) if
the cyclic order of the terminals is $s_1,\ldots,s_k,t_k,\ldots,t_1$ (by
reduction to the max-flow problem, after replacing each vertex by two
vertices connected by an arc, so that the problem is to find arc-disjoint
paths in this new graph)~\cite{hp-lbdpp-02}.  Our goal is to solve the
vertex-disjoint paths problem with minimal total length in
case~(\ref{enum_prob_i}).  We give an algorithm to do this in $O(kn\log n)$
time (see Figure~\ref{fig_problem}):
\begin{theorem}\label{thm_main}
  Let $G$ be a planar directed graph with $n$ vertices and arcs, each arc
  having a nonnegative length.  Let $s$ and $t$ be two distinct faces
  of~$G$; let $s_1,\ldots,s_k$ be vertices incident with~$s$; let
  $t_1,\ldots,t_k$ be vertices incident with~$t$.  Then we can compute $k$
  pairwise vertex-disjoint paths connecting the pairs $(s_i,t_i)$ in~$G$,
  with minimal total length, in $O(kn\log n)$ time.
\end{theorem}
The value of $k$ is not fixed in this result.  Note that this theorem also
holds if~$G$ is an undirected graph: simply replace every edge of this
graph by two oppositely directed arcs and apply the previous result to this
new graph.  The same problem for \emph{non-crossing} shortest paths, that
is, paths that are allowed to overlap along vertices and edges but not to
cross in the plane, is solvable in $O(n\log n)$ time, as shown by Takahashi
et al.~\cite{tsn-snppg-96}.

The high-level approach of our algorithm is the following.  We first show
that we may assume without loss of generality that~$G$ satisfies some
additional properties and transform~$G$ into another planar directed
graph~$D$; in this graph, it suffices to solve the same problem for
\emph{arc-disjoint} instead of \emph{vertex-disjoint} paths
(Section~\ref{sec_prelim}).  Then we translate our problem in terms of
flows in the graph~$D$ (Section~\ref{sec_flows}).  In
Section~\ref{sec_resid}, we introduce the residual graph and state some of
its properties that we will use.  In Section~\ref{sec_increase}, we explain
how to increase the value of an integer flow.  By repeated applications of
this algorithm, we obtain vertex-disjoint paths in~$G$ between the
terminals, but they may fail to connect the pairs $(s_i,t_i)$.  We show
that it suffices to ``rotate'' the flow a few times to change the
connections between the terminals (Section~\ref{sec_winding}) and explain
how to do that efficiently (Section~\ref{sec_turn}).  A generalization of
the notion of potential allows us to assume that all lengths in the
residual graph are nonnegative, which makes the algorithm efficient.

\section{Preliminaries}\label{sec_prelim}

We assume that we are given an embedding of the directed graph~$G$ in the
plane.  More precisely, only a \emph{combinatorial embedding} of $G$ is
necessary, which means that the cyclic order of the arcs around a vertex is
known.

We can assume that~$G$ is connected and that $t$ is the outer face of the
embedding of $G$.  Up to re-indexing the pairs $(s_i,t_i)$, we may assume
that $s_1,\ldots,s_k$ and $t_1,\ldots,t_k$ are in clockwise order: indeed,
if such a reordering does not exist, then there cannot exist
vertex-disjoint paths connecting the pairs $(s_i,t_i)$.

We may assume that each terminal vertex has degree one as follows: to each
terminal vertex $s_i$ (resp.\ $t_i$), attach an arc (of length zero, for
example) $(s'_i,s_i)$ (resp.\ $(t_i,t'_i)$) inside $s$ (resp.\ $t$), where
$s'_i$
(resp.\ $t'_i$) is a new vertex; use the $s'_i$ and the $t'_i$ as
terminals, instead of the $s_i$ and the $t_i$.  Clearly, any solution to
the problem in this augmented graph yields a solution in the original
graph~$G$.

\begin{figure}
  \centerline{%
    \psfrag{a1}{$s_1$}\psfrag{a2}{$s_2$}\psfrag{a3}{$s_3$}\psfrag{a4}{$s_4$}%
    \psfrag{b1}{$t_1$}\psfrag{b2}{$t_2$}\psfrag{b3}{$t_3$}\psfrag{b4}{$t_4$}%
    \includegraphics[width=.35\linewidth]{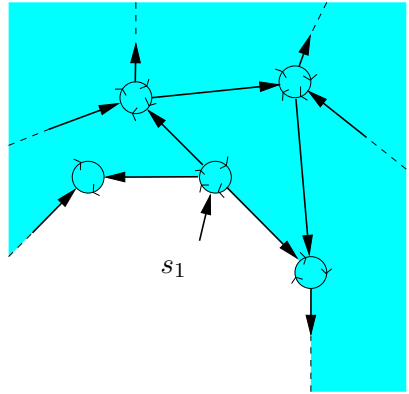}}
  \caption{Construction of the graph $D=(W,A)$ from the graph~$G$.  The
    thin arcs on the rings have length zero.}
  \label{fig_graphd}
\end{figure}

We transform $G$ into another directed planar graph $D=(W,A)$ by replacing
each non-terminal vertex $v$ of~$G$ by a small clockwise ``ring'' of arcs;
see Figure~\ref{fig_graphd}.  Every arc $a$ of~$D$ that is on no ring
corresponds to an arc of~$G$ and its length, $\lambda(a)$, is the length of
this arc in~$G$; it is thus nonnegative.  The length $\lambda(a)$ of an
arc~$a$ on a ring is zero.  The function $\lambda$ is fixed in this whole
paper.

An \emph{$(s,t)$-path} in~$D$ or~$G$ is a path from some vertex
in~$\{s_1,\ldots,s_k\}$ to some vertex in~$\{t_1,\ldots,t_k\}$; an
\emph{$(s_i,t_i)$-path} is a path connecting some pair of terminals
$(s_i,t_i)$.

\begin{proposition}\label{prp_gd}
  Let $P$ be a minimum-length set of $k$ vertex-disjoint $(s_i,t_i)$-paths
  in~$D$.  Then $P$ gives, in $O(n)$ time, a minimum-length set of $k$
  vertex-disjoint $(s_i,t_i)$-paths in~$G$.  If no such set~$P$ exists,
  then the original problem in~$G$ has no solution.
\end{proposition}
\proof
  Consider such a set of $(s_i,t_i)$-paths~$P$ in~$D$.  We claim that a
  given ring $r$ of~$D$ can be used by at most one path in~$P$.  Indeed,
  since $s$ and~$t$ are distinct faces, $\RR^2\setminus\{s\cup t\}$ is an
  annulus.  Since the paths in~$P$ are vertex-disjoint and connect $s$
  to~$t$, every point of the annulus that does not belong to a path in~$P$
  is on the left of exactly one path and on the right of exactly one path
  in~$P$.  In particular, the center~$c$ of~$r$ is on the right of exactly
  one path in~$P$.  But every path using~$r$ has~$c$ on its right, because
  the arcs of~$r$ are oriented clockwise.  This proves the claim.

  Thus, $P$ corresponds, in~$G$, to $k$ pairwise vertex-disjoint
  $(s_i,t_i)$-walks.  Removing the loops from these walks in $O(n)$ time
  does not increase the total length and gives a set of $k$ vertex-disjoint
  $(s_i,t_i)$-paths in~$G$.

  Conversely, any solution of the original vertex-disjoint problem in~$G$
  gives a set of $k$ vertex-disjoint paths in~$D$, of the same length,
  connecting the appropriate pairs of terminals.  So the paths obtained in
  the previous paragraph have minimal total length; furthermore, if no such
  set of paths~$P$ exists, then the problem in~$G$ admits no solution.
\qed

So we reduced the problem in~$G$ to the same problem in the graph~$D$.  The
point now is that the vertices of~$D$ have degree three, except the
terminals, which have degree one; because of these degree conditions, a set
of \emph{arc}-disjoint $(s,t)$-paths or circuits in~$D$ is actually a set
of \emph{vertex}-disjoint $(s,t)$-paths or circuits in~$D$, so we now have
to solve a problem on \emph{arc}-disjoint paths.  This enables a flow
approach on~$D$, which we will develop in the next section.

\section{Flows and winding numbers}\label{sec_flows}

In this paper, a \emph{flow} in $D=(W,A)$ is an element $x\in\RR^A$ such
that:
\begin{enumerate}
\item for each arc $a\in A$, $0\leq x(a)\leq 1$;
\item for each non-terminal vertex $v$, the following \emph{flow
    conservation law} holds:
  \[
  \sum_{a\,|\,v=\source(a)}x(a)=\sum_{a\,|\,v=\target(a)}x(a).
  \label{eq_conserv}
  \]
\end{enumerate}
The \emph{value} of a flow $x$ equals the total flow leaving the vertices
$s_1,\ldots,s_k$: if $a_i$ is the arc incident with~$s_i$, then the value
of~$x$ equals $\sum_{i=1}^kx(a_i)$.  A \emph{circulation} is a flow of
value zero.  A \emph{length function} (or \emph{cost function}) $\kappa$
on~$D$ is an element of~$\RR^A$; $\lambda$ is a length function.  The
\emph{length} (or \emph{cost}) of a flow $x$ with respect to~$\kappa$ is
defined to be $\kappa\transp x$.

An \emph{integer} flow is a flow in $\{0,1\}^A$; it is a set of
arc-disjoint $(s,t)$-paths and circuits in~$D$.  Actually, by the degree
conditions on~$D$, it is a set of \emph{vertex-disjoint} $(s,t)$-paths and
circuits.

Let $A\inv$ be the set of arcs in~$A$ with reverse orientation.  If
$\kappa\in\RR^A$ is a length function, we define the length of an arc
$a\inv\in A\inv$ to be $\kappa(a\inv)=-\kappa(a)$.

Let $X\in\RR^{A\cup A\inv}$; we define $z^X\in\RR^A$ by
$z^X(a)=X(a)-X(a\inv)$.  If $\gamma$ is a walk in $(W,A\cup A\inv)$, by a
slight abuse of notation, we define $z^\gamma$ to be $z^X$, where $X(a)$
(resp.\ $X(a\inv)$) is the number of times $\gamma$ travels through the arc
$a$ (resp.\ $a\inv$).  The length of~$\gamma$ with respect to a length
function~$\kappa$ is thus $\kappa\transp z^\gamma$.

\begin{figure}
  \centerline{%
    \psfrag{a1}{$s_1$}\psfrag{a2}{$s_2$}\psfrag{a3}{$s_3$}\psfrag{a4}{$s_4$}%
    \psfrag{b1}{$t_1$}\psfrag{b2}{$t_2$}\psfrag{b3}{$t_3$}\psfrag{b4}{$t_4$}%
    \psfrag{p}{{\footnotesize $+1$}}\psfrag{m}{{\footnotesize $-1$}}%
    \psfrag{U}{$U$}
    \includegraphics[width=.35\linewidth]{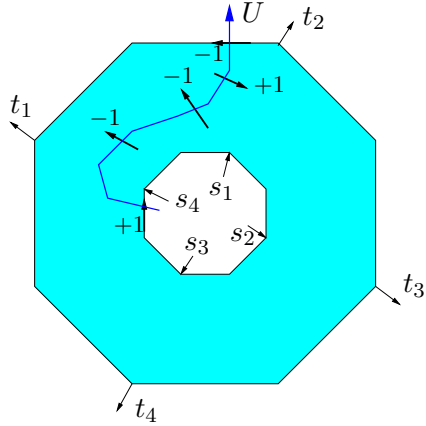}}
  \caption{The path $U$ in the dual graph~$D^*$ and the corresponding value
    of $u$ on the arcs of~$D$.  Only the non-zero values of~$u$ are
    indicated, on the arcs in bold lines.  Here $m=m_s-m_t=3-1=2$.}
  \label{fig_winding}
\end{figure}

We now want to take into account how a flow ``turns around'' the inner
face~$s$ of $G$.  To do this, consider the (undirected) dual graph $D^*$
of~$D$, that is, the planar graph that has one vertex $f^*$ inside each
face $f$ of~$D$ and such that $f_1^*$ and $f_2^*$ are connected by an edge
$e^*$ if and only if $f_1$ and $f_2$ are separated by an arc $e$ in $D$; in
that case, $e^*$ crosses $e$ but no other arc of~$D$.
Let $U$ be a path (fixed in this whole paper) from $s^*$ to~$t^*$ in~$D^*$
(Figure~\ref{fig_winding}).  For each arc~$a$ in $A$, define $u(a)$ to be
$0$ if $a$ does not cross~$U$, $+1$ if $a$ crosses $U$ from left to right,
and $-1$ if $a$ crosses $U$ from right to left.  This defines an element
$u\in\RR^A$.  The \emph{winding number} of a flow~$x$ equals $u\transp x$,
the value of the flow through~$u$ counted algebraically.  Also, for any
$X\in\RR^{A\cup A\inv}$, the winding number of~$X$ is $u\transp z^X$.

Let $m_s\in[1,k]$ be such that the first arc of~$U$ is, in the cyclic
order around the face~$s$, between $s_{m_s}$ and $s_{m_s+1\bmod k}$.
Similarly, let $m_t$ be such that the last arc of~$U$ is between $t_{m_t}$
and $t_{m_t+1\bmod k}$.  Let $m=m_s-m_t$.

The following lemma will be used repeatedly.
\begin{lemma}\label{lem_planar}
  Let $\gamma$ be any circuit in $(W,A\cup A\inv)$.  Then the winding
  number of~$\gamma$ belongs to $\{-1,0,+1\}$.  If~$\gamma$ encloses~$s$ in
  the plane, then it has winding number~$+1$ if it is clockwise and~$-1$ if
  it is counter-clockwise.  Otherwise, $\gamma$ has winding number $0$.
\end{lemma}
\proof
  This is a consequence of the Jordan curve theorem.  The winding number
  of~$\gamma$ is the number of times the path~$U$ crosses~$\gamma$ from the
  right to the left, minus the number of times $U$ crosses~$\gamma$ from
  the left to the right.  Assume $\gamma$ is clockwise, the other case
  being analogous.  The winding number of~$\gamma$ is the number of times
  $U$ exits the region enclosed by~$\gamma$ minus the number of times it
  enters this region.

  If $\gamma$ does not enclose~$s$, then both endpoints of~$U$ are
  outside~$\gamma$, so the winding number is zero.  If $\gamma$
  encloses~$s$, the source of~$U$ is inside the region enclosed by~$\gamma$
  while its target is outside, so the winding number is~$+1$.
\qed

We can now reformulate our arc-disjoint paths problem in~$D$ in terms of
flows in~$D$:
\begin{proposition}\label{prp_corr}
  Let~$x$ be an integer flow in~$D$ of value~$k$ with minimal cost subject
  to the condition that its winding number, modulo~$k$, equals $m$.  Then $x$
  gives, in $O(n)$ time, $k$ vertex-disjoint $(s_i,t_i)$-paths in~$D$ of
  minimal total length.  If there exists no such flow, then there does not
  exist $k$ vertex-disjoint $(s_i,t_i)$-paths in~$D$.
\end{proposition}
\proof
  As noted above, the degree conditions on~$D$ imply that
  the flow~$x$ is a set of \emph{vertex}-disjoint $(s,t)$-paths or
  circuits in~$D$.  Let $\gamma$ be a circuit in~$x$.  If~$\gamma$ has
  non-zero winding number, then $\gamma$ separates $s$ and~$t$, which
  implies that $x$ has value zero, a contradiction.  If~$\gamma$ has
  winding number zero, then removing it from~$x$ yields another flow with
  the same properties.  Since we can remove such circuits in $O(n)$ time,
  we may assume that $x$ contains only $(s,t)$-paths.  By the assumption on
  the winding number, these paths connect the pairs $(s_i,t_i)$, for
  $i=1,\ldots,k$.

  Furthermore, any $k$ vertex-disjoint $(s_i,t_i)$-paths in~$D$ correspond
  to a flow in~$D$ of value~$k$ and of winding number equal, modulo~$k$,
  to~$m$.  It follows that the paths obtained have minimal total length.
\qed

By Propositions \ref{prp_gd} and~\ref{prp_corr}, to prove
Theorem~\ref{thm_main}, it suffices to show that we can, in $O(kn\log n)$
time, find an integer flow in~$D$ of value~$k$ and with minimal cost
subject to the condition that its winding number, modulo~$k$, equals $m$.

\section{The residual graph}\label{sec_resid}

In this section, we introduce the \emph{residual graph} of $D$ in the
special case of integer flows; it is a classical tool for dealing with
maximal flows and flows of minimal cost~\cite[Chapters~10--12]{s-cope-03}.

Let $x$ be an integer flow on $D=(W,A)$.  Let $A_x$ be the subset of $A\cup
A\inv$ defined by
\[
A_x=\{a\ |\ x(a)=0\} \cup \{a\inv\ |\ x(a)=1\}.
\]
The \emph{residual graph} of $D$ with respect to $x$ is the directed graph
$D_x=(W,A_x)$; it is thus the graph obtained from~$D$ by reversing the sign
of the length and winding number and the orientation of the arcs $a$ such
that $x(a)=1$.

The following lemma explains the interest of the residual graph; the first
two assertions are well-known.

\begin{lemma}\label{lem_resid}
  Let $x$ be an integer flow in~$D$.
  \begin{enumerate}\renewcommand{\theenumi}{\roman{enumi}}
  \item\label{enum_resid_i} $D_x$ has no $(s,t)$-path if and only if $x$
    has maximal value in~$D$ among all flows.
  \item\label{enum_resid_ii} Assume that $x$ has maximal value in~$D$; let
    $\kappa$ be a length function.  Then $D_x$ has no negative-length
    directed circuit with respect to~$\kappa$ if and only if $x$ has
    minimal cost, with respect to~$\kappa$, among all flows in~$D$ with the
    same value.
  \item\label{enum_resid_iii} Assume $x$ has maximal value in~$D$.  Then $D_x$
    has no directed circuit with winding number one if and only if $x$ has
    maximal winding number among all flows in~$D$ with the same value.
  \end{enumerate}
\end{lemma}
\proof
  In these three assertions, the ``if'' part is easy: If $D_x$ has an
  $(s,t)$-path or circuit $\gamma$, then, by construction of~$D_x$,
  $y:=x+z^\gamma$ is an integer flow in~$D$; its cost equals the cost
  of~$x$ in~$D$ plus the cost of~$\gamma$ in~$D_x$; its winding number
  equals the winding number of~$x$ plus the winding number of~$\gamma$; and
  its value equals the value of~$x$ plus one if $\gamma$ is a path, or the
  value of~$x$ if $\gamma$ is a circuit.

  Conversely, let $x$ be an integer flow in~$D$ and let $y$ be any flow
  in~$D$.  Consider $y-x$ in the graph~$D$.  By construction of~$D_x$, this
  is a flow in~$D_x$, in the sense that the flow conservation law holds at
  each vertex of~$D$ (except at the terminals) and that, for each arc $a\in
  A$, we have $(y-x)(a)\geq 0$ if $a\in A_x$ and $(y-x)(a)\leq 0$ if
  $a\inv\in A_x$.  In particular, $y-x$ can be written as $\sum_{\gamma\in
    Z}\alpha_\gamma z^\gamma$, where $Z$ is a set of $(s,t)$-paths,
  $(t,s)$-paths, and circuits in~$D_x$, and the $\alpha_\gamma$ are
  positive real numbers.

  Now, to prove the ``only if'' part of~(\ref{enum_resid_i}), simply note
  that, if $D_x$ has no $(s,t)$-path, then there is no $(s,t)$-path in~$Z$;
  thus, the value of $y$ cannot be greater than the value of~$x$.  To prove
  the ``only if'' part of (\ref{enum_resid_ii}) and~(\ref{enum_resid_iii}),
  assume that $x$ and~$y$ both have maximal value in~$D$.  Then,
  by~(\ref{enum_resid_i}), $Z$ contains no $(s,t)$-path, hence also no
  $(t,s)$-path, hence only circuits.  If $D_x$ has no negative-length
  directed circuit, the cost of $y$ is at least the cost of $x$; this
  proves~(\ref{enum_resid_ii}).  If $D_x$ has no directed circuit with
  winding number one, then $y$ cannot have winding number higher than~$x$,
  for otherwise $y-x$ would contain at least one circuit with positive
  winding number, hence with winding number one (Lemma~\ref{lem_planar}).
  This proves~(\ref{enum_resid_iii}).
\qed

A length function $\kappa$ is \emph{nonnegative on~$D_x$} if $\kappa$ is
nonnegative on every arc in~$A_x$; that is, for each $a\in A$,
$\kappa(a)\geq 0$ if $x(a)=0$ and $\kappa(a)\leq 0$ if $x(a)=1$.

\section{Increasing the flow in~$D$}\label{sec_increase}

In this section, we explain how to compute a minimum-cost flow in~$D$ in
$O(kn\log n)$ time.  The algorithm uses only very classical minimum-cost
flow techniques, but we indicate it for completeness and because
Section~\ref{sec_turn} will use some similar ideas.

Let $p\in\ZZ$.  A \emph{$p$-flow} is an \emph{integer} flow in~$D$ of value~$p$.
Let $\kappa$ and $\kappa'$ be two length functions on~$D$; we write
$\kappa\simeq\kappa'$ if $\kappa\transp z^\gamma=\kappa\ptransp z^\gamma$ for
each closed walk $\gamma$ in~$(W,A\cup A\inv)$.  (This notion is equivalent
to the notion of potential.)

\begin{lemma}\label{lem_sim}
  Let $\kappa\simeq\kappa'$.  Then any minimum-cost $k$-flow with respect
  to~$\kappa$ is also a minimum-cost $k$-flow with respect to~$\kappa'$.
\end{lemma}
\proof
  By Lemma~\ref{lem_resid}(\ref{enum_resid_ii}), a $k$-flow~$x$ has minimum
  cost with respect to~$\kappa$ if and only if $D_x$ has no negative-length
  circuit with respect to~$\kappa$.  Since $\kappa\simeq\kappa'$, circuits
  in~$D_x$ have the same length with respect to~$\kappa$ and to~$\kappa'$.
\qed

The following result follows from classical minimum-cost flow techniques.
\begin{lemma}\label{lem_path}
  Let $x$ be a $p$-flow in~$D$ and let $\kappa$ be a length function that
  is nonnegative on~$D_x$.  Then, in $O(n\log n)$ time, we can find a
  $(p+1)$-flow~$x'$ and a length function $\kappa'\simeq\kappa$ that is
  nonnegative on~$D_{x'}$, unless $x$ has maximal value.
\end{lemma}
\proof
  We temporarily add to~$D_x$ two vertices $s$ and $t$, and arcs $(s,s_i)$
  and $(t_i,t)$ of length zero, for $i=1,\ldots,k$.  Let $D'_x$ be the
  resulting graph.  We compute a shortest path tree of~$D'_x$ with
  root~$s$, with respect to~$\kappa$, in $O(n\log n)$ time using Dijkstra's
  algorithm~\cite{d-ntpcg-59} speeded up with Fibonacci
  heaps~\cite{ft-fhuin-87}, because all lengths are nonnegative%
  \footnote{We could do that in $O(n)$ time using the algorithm by
    Henzinger et al.~\cite{hkrs-fspap-97}, but that would not change the
    asymptotic complexity of the entire algorithm.}%
  .  If there is no path from $s$ to~$t$ in~$D'_x$, then $D_x$ has no
  $(s,t)$-path, hence, by Lemma~\ref{lem_resid}(\ref{enum_resid_i}), $x$
  has maximal value.

  Otherwise, for each vertex~$v$ of~$D'_x$, let $d(v)$ be the distance from
  $s$ to~$v$ with respect to $\kappa$, as computed by Dijkstra's algorithm
  above.  For each arc $a=(u,v)$ of~$A_x$, we have $d(v)\leq
  d(u)+\kappa(a)$ by the triangle inequality, with equality if $a$ is
  on the shortest path tree.  For each arc $a=(u,v)$ of~$A_x$, let
  $\kappa'(a)=\kappa(a)+d(u)-d(v)$; clearly, $\kappa'\simeq\kappa$.  We
  have $\kappa'(a)\geq 0$, and $\kappa'(a)=0$ if $a$ is on the
  shortest path tree.  Let $\gamma$ be the $(s,t)$-path in~$D_x$
  corresponding to the path from $s$ to~$t$ in~$D'_x$ in the shortest path
  tree.  Now, let $x'=x+z^\gamma$; since $\kappa'$ is nonnegative on the
  arcs of~$D_x$ and is zero on the arcs of~$\gamma$, it is nonnegative
  on~$D_{x'}$.
\qed

Starting with the zero flow~$x$ (for which $D_x=D$) and the length
function~$\kappa=\lambda$,
we repeatedly apply Lemma~\ref{lem_path}.  We obtain a flow~$x_0$ with
maximal value~$p$ and a length function~$\kappa_0\simeq\lambda$ such that
$\kappa_0$ is nonnegative on~$D_{x_0}$.  This takes $O(pn\log n)=O(kn\log
n)$ time.  If~$p<k$, then the original problem has no solution, hence we
stop here.  Otherwise, Lemmas \ref{lem_resid}(\ref{enum_resid_i})
and~\ref{lem_sim} imply that $x_0$ is a minimum-cost $k$-flow with respect
to~$\lambda$.  Let $w_0$ be the winding number of~$x_0$.  If $w_0\equiv
m\pmod k$, then we are done by Propositions \ref{prp_gd}
and~\ref{prp_corr}; so we henceforth assume $w_0\not\equiv m\pmod k$.

\section{Finding the winding number}\label{sec_winding}

A \emph{$(k,w)$-flow} is an integer flow in~$D$ of value~$k$ and winding
number~$w$.  Let $w_1$ and $w_2$ be the integers equal, modulo~$k$, to~$m$ 
that are the closest to~$w_0$ and satisfy $w_1<w_0<w_2$.  The following
proposition states that the problem boils down to finding minimum-cost
$(k,w)$-flows, for $w=w_1$ and $w=w_2$:

\begin{proposition}\label{prp_twoposs}
  There is a minimum-cost integer flow in~$D$ (with respect to~$\lambda$)
  of value~$k$ and winding number equal, modulo~$k$, to~$m$ that is either a
  $(k,w_1)$-flow or a $(k,w_2)$-flow.
\end{proposition}
\proof
  For every integer~$w$, let $\mu_w$ be the minimal cost of the
  $(k,w)$-flows.  (It is infinite if no $(k,w)$-flow exists.)  By
  Lemma~\ref{lem_resid}(\ref{enum_resid_iii}), the set $\{w\ |\
  \mu_w<\infty\}$ is an interval of integers.

  We show that for every integer~$w$ such that $\mu_{w-1}$, $\mu_w$, and
  $\mu_{w+1}$ are finite, we have
  \begin{equation}
    2\mu_w\leq \mu_{w-1}+\mu_{w+1}.
    \label{eq_convex}
  \end{equation}
  Indeed, let $x$ and~$x'$ be minimum-cost $(k,w-1)$- and $(k,w+1)$-flows,
  respectively.  Then $x'-x$ gives a nonnegative integer circulation
  in~$D_x$ of winding number~$2$, i.e., a flow $y$ of value zero in~$D$
  such that, for each $a\in A$, $y(a)\geq 0$ if $a\in A_x$ and $y(a)\leq 0$
  if $a\inv\in A_x$.  So the support of~$x'-x$ contains a directed
  circuit~$\gamma$ in~$D_x$ of positive winding number, hence~$1$.  Then
  $x+z^\gamma$ and~$x'-z^\gamma$ are both $(k,w)$-flows.  Thus
\[
  2\mu_w \leq \lambda\transp(x+z^\gamma)+\lambda\transp(x'-z^\gamma)
  = \lambda\transp x+\lambda\transp x'
  = \mu_{w-1}+\mu_{w+1},
\]
  which proves~(\ref{eq_convex}).

  So $\mu_w$ is monotonically non-increasing for $w\leq w_0$ and
  monotonically non-decreasing for $w\geq w_0$.  Thus
  Proposition~\ref{prp_twoposs} holds.
\qed

\section{Rotating the flow in~$D$}\label{sec_turn}

Let $\kappa$ and $\kappa'$ be two length functions on~$D$; we write
$\kappa\sim\kappa'$ if $\kappa\transp z^\gamma=\kappa\ptransp z^\gamma$
for each closed walk~$\gamma$ \emph{with winding number zero} in~$(W,A\cup
A\inv)$.  Clearly, $\kappa\simeq\kappa'$ implies $\kappa\sim\kappa'$.

\begin{proposition}\label{prp_simeq}
  Let $\kappa\sim\kappa'$.  Then any minimum-cost $(k,w)$-flow with
  respect to~$\kappa$ is also a minimum-cost $(k,w)$-flow with respect
  to~$\kappa'$.
\end{proposition}
\proof
  Let $x$ and~$y$ be two $(k,w)$-flows in~$D$.  Then $y-x$ is a circulation
  in~$(W,A\cup A\inv)$, i.e., a sum of terms of the form~$z^\gamma$,
  where~$\gamma$ is a circuit in~$(W,A\cup A\inv)$.  Furthermore, there are
  as many circuits with winding number~$+1$ as with winding number~$-1$ in
  this sum.

  We have $(\kappa'-\kappa)\transp z^\gamma=0$ for every such circuit with
  winding number zero.  Moreover, if~$\gamma$ has winding number~$+1$
  and~$\gamma'$ has winding number~$-1$, it follows from the definition
  of~``$\sim$'' that
  $\kappa\transp(z^\gamma+z^{\gamma'})=\kappa\ptransp(z^\gamma+z^{\gamma'})$.
  We thus have $\kappa\transp(y-x)=\kappa\ptransp(y-x)$, implying the
  result.
\qed

We view $D$ as an undirected planar graph~$H$; $s$ and~$t$ are two faces
of~$H$.  Let $H^*$ be its dual graph.  If $e$ is an \emph{oriented} edge
of~$H$, then $e^*$ is the dual edge oriented so that $e^*$ crosses $e$ from
right to left.

\begin{figure}
  \centerline{%
    \psfrag{s}{$s$}\psfrag{t}{$t$}
    \includegraphics[width=.35\linewidth]{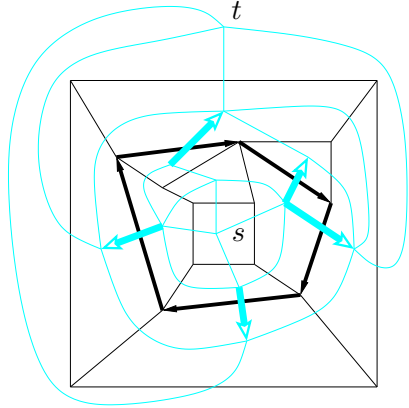}}
  \caption{Illustration of Lemma~\ref{lem_circuit_cut}: A minimal cut
    in~$H^*$ corresponds to a circuit with winding number one in~$H$.  The
    primal graph~$H$ is depicted in black lines, with thicker lines for the
    arcs of the circuit.  The dual graph~$H^*$ is depicted in light color,
    with thicker lines for the arcs of the cut.}
  \label{fig_cut}
\end{figure}

A \emph{cut} of~$H^*$ is a set $X^*$ of \emph{oriented} edges of~$H^*$ such
that any directed path from $s^*$ to~$t^*$ uses at least one oriented edge
of~$X^*$.  The following lemma is inspired by Reif~\cite[Propositions 1
and~2]{r-mscpu-83}.  See Figure~\ref{fig_cut}.

\begin{lemma}\label{lem_circuit_cut}
  Let $X$ be a set of oriented edges of~$H$.  Then $X$ contains the
  oriented edges of some circuit with winding number one in~$H$ if and only
  if $X^*$ is a cut of~$H^*$.
\end{lemma}
\proof
  If we have a directed circuit $\gamma$ with winding number one, then its
  dual is a cut.  Indeed, consider an $(s^*,t^*)$-path $\pi$ in~$H^*$.  The
  face $s$ belongs to the interior of~$\gamma$, while the face~$t$ belongs
  to the exterior of~$\gamma$; let $e^*$ be the first oriented edge
  of~$\pi$ that crosses~$\gamma$; its source is inside~$\gamma$ while its
  target is outside~$\gamma$.  By our choice of orientation, $e$ belongs
  to~$\gamma$.

  Conversely, let $X^*$ be a cut of~$H^*$; we will prove that $X$ contains
  a circuit with winding number one.  Without loss of generality, we may
  assume that $X^*$ is a cut that is \emph{minimal} with respect to
  inclusion.

  First, label ``S'' a face $f$ of~$H$ if there is, in~$H^*$, a path from
  $s^*$ to~$f^*$ that does not use any oriented edge of~$X^*$.  Similarly,
  label ``T'' a face $f$ of~$H$ if there is, in~$H^*$, a path from $f^*$
  to~$t^*$ that does not use any oriented edge of~$X^*$.  Since $X^*$ is a
  cut, no face of~$H$ is labeled both ``S'' and ``T''.  We claim that $X$
  is precisely the set of oriented edges of~$H$ whose right face is labeled
  ``S'' and whose left face is labeled ``T''.  Clearly, such edges must
  belong to~$X$.  Conversely, let $e$ be an oriented edge of~$X$; by
  minimality of~$X$, there is an $(s^*,t^*)$-path in~$H^*$ that avoids
  $(X\setminus e)^*$ and uses $e^*$ exactly once.  Thus the source of~$e^*$
  is reachable from~$s^*$ without using any oriented edge of~$X^*$, and
  $t^*$ is reachable from the target of~$e^*$ without using any oriented
  edge of~$X^*$.  This proves the claim.  In particular, every face of~$H$
  is labeled either ``S'' or ``T''.

  Let $S$ be the subset of the plane made of the faces labeled ``S'',
  together with the \emph{open} edges whose both incident faces are labeled
  ``S''.  Similarly, let $T$ be the union of the faces labeled ``T''
  together with the open edges whose both incident faces are labeled ``T''.
  By the previous paragraph, $S$ and $T$ are disjoint subsets of the plane,
  and they are connected.  Let $v$ be a vertex of~$H$.  We claim that there
  cannot be four faces incident with~$v$, in this cyclic order around~$v$,
  that belong respectively to $S$, $T$, $S$, and~$T$.  This follows from
  the Jordan curve theorem: assume that we have such faces.  Then, by
  connectivity of~$S$, there is a simple closed curve in~$S\cup\{v\}$ that
  goes through~$v$ and has faces of~$T$ on both sides of it at~$v$.  This
  curve does not intersect~$T$ and separates~$T$, contradicting its
  connectivity.

  The two previous paragraphs together imply that either $X$ has no edge
  incident with~$v$, or $X$ has exactly one oriented edge whose target is~$v$
  and one oriented edge whose source is~$v$.  Thus $X$ is a union of
  vertex-disjoint circuits.  Let $\gamma$ be such a circuit; since $S$
  and~$T$ are connected, and since the faces on the left (resp.\ right)
  of~$\gamma$ are in~$T$ (resp.~$S$), $\gamma$ has winding number one.
  Hence $X$ contains a circuit with winding number one.
\qed

\begin{proposition}\label{prp_turn}
  Let $x$ be a $(k,w)$-flow in~$D$ and let $\kappa$ be a length function
  that is nonnegative on~$D_x$.  Then, in $O(n\log n)$ time, we can find a
  $(k,w+1)$-flow~$x'$ and a length function $\kappa'\sim\kappa$ that is
  nonnegative on~$D_{x'}$, unless there is no $(k,w')$-flow with $w'>w$.
\end{proposition}
\proof
  Let~$e$ be an oriented edge of~$H$; if $e$ corresponds to an arc~$a$
  of~$A_x$, then we define the length of~$e$ in~$H$ to be $\kappa(a)\geq
  0$; otherwise, we define the length of~$e$ to be~$\infty$.  So a walk
  in~$D_x$ corresponds to a walk in~$H$ of the same length, and a walk
  in~$H$ corresponds to a walk in~$D_x$ if and only if it has finite
  length.  Define the \emph{capacity} $c(e^*)$ of an oriented edge $e^*$
  of~$H^*$ to be the length of~$e$.

  We can detect in $O(n)$ time whether the oriented edges of finite capacity
  constitute a cut in~$H^*$.  If this is not the case, then every cut must
  use an oriented edge of infinite capacity, hence, by
  Lemma~\ref{lem_circuit_cut}, $D_x$ has no circuit of winding number one.
  It follows that $x$ has maximal winding number among all $k$-flows, by
  Lemma~\ref{lem_resid}(\ref{enum_resid_iii}).  Otherwise, we compute a
  minimal cut in~$H^*$, which corresponds to a shortest circuit with
  winding number one in~$D_x$, as follows.

  A \emph{flow} in~$H^*$ is a function~$\varphi$ that associates, to each
  oriented edge~$e^*$ of~$H^*$, a real number that is nonnegative and no
  greater than~$c(e^*)$, such that the flow conservation law holds at each
  vertex of~$H^*$ except at $s^*$ and $t^*$.  The \emph{value} of~$\varphi$
  is the total flow leaving $s^*$.

  In $O(n\log n)$ time, we compute a flow $\varphi$ of maximal value in
  $H^*$ with respect to these capacities, using the algorithm by Borradaile
  and Klein~\cite{bk-namsd-06}.  It is well-known, by the ``max-flow
  min-cut'' theorem~\cite[Theorem 10.3]{s-cope-03}, that $\varphi$
  corresponds to a cut of minimal cost in~$H^*$: the cut is the set of
  oriented edges that leave the set of vertices reachable from~$s^*$ by
  using only oriented edges~$e^*$ of~$H^*$ such that $\varphi(e^*)<c(e^*)$
  or $\varphi(e^{*-1})>0$.

  Such a cut~$X^*$ can be computed in $O(n)$ time.  Moreover, by replacing
  all the zero capacities in~$H^*$ by infinitesimally small capacities
  before applying the maximal flow algorithm, we may assume that~$X^*$ is a
  cut that is minimal with respect to inclusion.  By
  Lemma~\ref{lem_circuit_cut}, we thus obtain a circuit~$\gamma$ of winding
  number one that has minimal length in~$D_x$.

  For each arc~$a$ of~$A\cup A\inv$, let
  $\kappa'(a)=\kappa(a)-\varphi(a^*)+\varphi(a^{*-1})$; we have
  $\kappa'(a)=-\kappa'(a\inv)$, hence this defines a length function.  If
  $a\in A_x$, we have $\varphi(a^*)\leq\kappa(a)$, so $\kappa'(a)\geq0$.
  If $a$ belongs to~$\gamma$, we have $\varphi(a^*)=\kappa(a)$ and
  $\varphi(a^{*-1})=0$, so $\kappa'(a)=0$.

  We claim that $\kappa'\sim\kappa$.  By the flow conservation law
  in~$H^*$, $\kappa'-\kappa$ is a linear combination of functions of the
  form~$z^\gamma$, where $\gamma^*$ is an $(s^*,t^*)$-path or a circuit
  in~$H^*$; so it suffices to prove that~$z^{\gamma\top}\delta=0$ for each
  closed walk~$\delta$ with winding number zero.  But
  $z^{\gamma\top}\delta$ equals the number of times $\delta$ crosses
  $\gamma^*$ from left to right minus the number of times $\delta$ crosses
  $\gamma^*$ from right to left.  This always equals zero if~$\gamma$ is a
  circuit; if~$\gamma$ is an $(s^*,t^*)$-path, this equals zero because
  $\delta$ has winding number zero (as in the proof of
  Lemma~\ref{lem_planar}).  This proves $\kappa'\sim\kappa$.

  Now, let $x'=x+z^\gamma$.  The length function $\kappa'$ is nonnegative
  on the arcs of~$D_x$ and is zero on the arcs of~$\gamma$, so it is
  nonnegative on~$D_{x'}$.
\qed

To conclude, recall that the $k$-flow~$x_0$ and the length
function~$\kappa_0$ have been computed in Section~\ref{sec_increase};
$\kappa_0\sim\lambda$ is nonnegative on~$D_{x_0}$; the integer $w_0$ is
the winding number of~$x_0$ and we have
\[
  w_0-k<w_1<w_0<w_2<w_0+k.
\]

Applying iteratively Proposition~\ref{prp_turn}, we can find a
$(k,w_2)$-flow $x_2$ and a length function $\kappa_2\sim\lambda$ that is
nonnegative on~$D_{x_2}$; thus, $x_2$ is a $(k,w_2)$-flow of minimal cost
with respect to~$\lambda$, by Lemma \ref{lem_resid}(\ref{enum_resid_ii})
and Proposition~\ref{prp_simeq}; if no such flow exists, we detect it
during the course of the algorithm.  Similarly, we can find a minimum-cost
$(k,w_1)$-flow.  This takes $O(kn\log n)$ time.  By Propositions
\ref{prp_twoposs}, \ref{prp_gd}, and~\ref{prp_corr}, the cheapest of these
two flows corresponds to the solution.  This concludes the proof of
Theorem~\ref{thm_main}.

\section*{Conclusion}

We have given an algorithm to compute minimum-length vertex-disjoint paths
connecting prescribed pairs $(s_i,t_i)$ of terminals in a planar graph,
where the $s_i$ and the $t_i$ are incident, respectively, with given faces
$s$ and~$t$.  The running time is $O(kn\log n)$, where $k$ is the number of
pairs of terminals and $n$ is the complexity of the graph.

We note that the techniques developed above allow to solve the same
problem, but fixing, in addition, the winding number of the set of paths
(or, equivalently, the homotopy classes of the paths in the annulus
$\RR^2\setminus\{s\cup t\}$).  This can be done by computing a minimum-cost
flow in the directed graph~$D$ and by rotating the flow until achieving the
correct winding number.  Since the absolute value of the winding number of
a flow is at most $n$, the complexity of the algorithm is $O(n^2\log n)$.

Finally, the result of this paper suggests some open questions.  How hard
is it to solve the minimum-length vertex-disjoint paths in
case~(\ref{enum_prob_ii}) of the introduction, namely, if all terminals lie
on the outer face (not necessarily in the order
$s_1,\ldots,s_k,t_k,\ldots,t_1$)?  And in the case where all the terminals
lie on two faces, but a path may have its two endpoints on the same face?
The problem extends to vertex-disjoint \emph{trees} whose leaves are fixed
on two faces of the graph (such trees, not necessarily of minimal length,
can be computed efficiently~\cite{san-fsfpg-90}).  Also, does our problem
remain polynomial-time solvable if each of the terminals has to be incident
with one of $p$ prescribed faces of the graph, if $p$ is fixed?  What about
the same problem for a graph embedded on a surface of fixed genus?

\section*{Acknowledgements}

We would like to thank Dion Gijswijt and G\"unter Rote for stimulating
discussions.

\end{document}